\documentclass{ifacconf}

\usepackage{amsmath}
\usepackage{graphicx}      % include this line if your document contains figures
\graphicspath{{figures/}}
\usepackage{natbib}        % required for bibliography

\newcommand{\norm}[1]{\left\lVert#1\right\rVert}
\newcommand\abs[1]{\left|#1\right|}

\DeclareMathOperator*{\argmin}{arg\,min}

\begin{document}
\begin{frontmatter}

\title{Analysis of risk levels for traffic on a multi-lane highway\thanksref{footnoteinfo}} 

\thanks[footnoteinfo]{This work has been supported by HE5386/13-15 and DAAD MIUR project.}

\author[MH]{Michael~Herty} 
\author[GV]{Giuseppe~Visconti} 

\address[MH]{Institut f\"{u}r Geometrie und Praktische Mathematik (IGPM),\\RWTH Aachen University, Aachen, Germany\\(e-mail: herty@igpm.rwth-aachen.de)}
\address[GV]{Institut f\"{u}r Geometrie und Praktische Mathematik (IGPM),\\RWTH Aachen University, Aachen, Germany\\(e-mail: visconti@igpm.rwth-aachen.de)}

\begin{abstract}                % Abstract of not more than 250 words.
	We present an analysis of risk levels on multi-lane roads. The aim is to use the crash metrics to understand which direction of the flow mainly influences the safety in traffic flow. In fact, on multi-lane highways interactions among vehicles occur also with lane changing and we show that they strongly affect the level of potential conflicts. In particular, in this study we consider the Time-To-Collision as risk metric and we use the experimental data collected on the A3 German highway.
\end{abstract}


\end{frontmatter}
%===============================================================================

\section{Introduction}

The analysis, the prediction and the control of critical traffic situations are important aspects of the modern world. In fact, with the increase of the number of circulating vehicles the risk of accidents is considerably increased. See the recent report~\cite{WHO}. Therefore, the necessity of forecasting e.g. the evolution of the flow as well as of the potential risks has arisen. These problems have motivated research both in the mathematical and in the engineering literature.

From the mathematical point of view, the analysis of traffic critical situations is carried out by means of mathematical models at different scales. Microscopic, macroscopic and kinetic simulations provide useful tools to study e.g. the links among traffic volumes and safety issues as well as the creation of shock or stop-and-go waves, which represent typical effects of instabilities and thus of potential risks in the flow. See e.g.~\cite{MoutariHertyEtAl,MoutariHerty,FregugliaTosin17}

In contrast, engineering studies on the risk levels in traffic flow are mainly based on a-posteriori descriptions of collected real data on roads. In this field experimental measurements are used to study the evolution of the safety indicators related to traffic. See e.g.~\cite{MinderhoudBovy,KuangEtAl,KuangQuYan,WangEtAl} Among these we recall the Headway Distribution, the Time-To-Collision or the Aggregated Crash Index.

In this paper we are aimed to take into account an intrinsic characteristic of traffic flow which is however usually neglected in the study of risk levels: the lane changing. We will focus in particular on the Time-To-Collision metric and we generalize the definition also for the direction orthogonal to the flow of vehicles. The Time-To-Collision indicator is defined as the remaining time until a collision between two vehicles would have occurred. For each pair of interacting vehicle, the complete formulation of the Time-To-Collision is given by
\begin{equation} \label{eq:extendedTTC}
	\text{TTC} = 
				 \begin{cases}
					- \frac{\Delta x}{\Delta v}, & \text{if $\Delta v < 0$, $\Delta a = 0$}\\[2ex]
					- \frac{\Delta v}{\Delta a} - \frac{\sqrt{\Delta v^2 - 2 \Delta x \Delta a}}{\Delta a}, & \text{if $\Delta v<0$, $\Delta a \neq 0$}\\[2ex]
					- \frac{\Delta v}{\Delta a} + \frac{\sqrt{\Delta v^2 - 2 \Delta x \Delta a}}{\Delta a}, & \text{if $\Delta v \geq 0$, $\Delta a < 0$}\\[2ex]
					+\infty, & \text{otherwise}. 
				 \end{cases}
\end{equation}
where $\Delta x$, $\Delta v$ and $\Delta a$ are the relative distance, speed and acceleration of the two vehicles. Here, we will study also the lateral safety by comparing it with the classical notion of the Time-To-Collision indicator concerning only the longitudinal driving tasks. The aim is to understand which direction of traffic flow influences mainly the safety level of a multi-lane road. This will be therefore an a-posteriori analysis based on experimental data collected on a German highway. At the same time we will give a macroscopic approximation of the Time-To-Collision indicator in order to show a link among the risk metric and the evolution in time of traffic volumes, giving also a hint on how macroscopic equations can be used for real-time predictions of safety measures.

In detail the paper is organized as follows. In Section~\ref{sec:dataset} we describe the German data-set recovering, from the microscopic quantities, the macroscopic data which lead to the fundamental diagrams of traffic. In Section~\ref{sec:TTC} we define the meaning of a car-following scenario on a multi-lane road and we define the Time-To-Collision risk indicators for the two directions of the flow. We further give a macroscopic approximation of the risk metric. In Section~\ref{sec:simulations} we show the analysis of the risk levels on the German highway comparing the result obtained by considering separately the flow along the road and across the lanes. Finally, we end in Section~\ref{sec:outlook} summarizing the result of the paper and giving an outlook for future research.

\section{Experimental data collection} \label{sec:dataset}

In this paper we consider a set of experimental data recorded on a German highway. We have two-dimensional trajectory data collected on a $80$~meters stretch of the westbound direction of the A3~highway near Aschaffenburg. Laser scanners detect the two-dimensional positions $P_i(t) = \left(x_i(t),y_i(t)\right)$ of each vehicle $i$ at time $t$ on the road segment with a temporal resolution of $0.2$~seconds for a total time of approximately $20$~minutes. Here the position $x$ is in driving direction, the position $y$ is across lanes. During the time observation, the laser scanners record the trajectories of $1290$~vehicles.

The road section consists of three lanes and an outgoing ramp. However, we only consider the stretch as if there is no ramp. In fact, the data show that the flow on the ramp does not influence the traffic conditions, namely the amount of traffic on the ramp is not significant. Taking into account only the three main lanes, the road width is $12$~meters.

The microscopic velocities of vehicles are recovered by the knowledge of their positions $P_i(t) = \left(x_i(t),y_i(t)\right)$ at each time. Since the road section is relatively short, we compute the velocity, both in $x$- and $y$-direction, of each vehicle by using a linear approximation in the least squares sense of $x_i(t)$ and $y_i(t)$, respectively. In other words we assume that the vehicle velocity is constant during the crossing of the road section and is exactly the slope of the linear fit. The maximum detected speed in $x$-direction is about $120$ kilometer per hour which means about $2.7$~seconds to travel the $80$~meters of the road section. Instead, the maximum detected speed in $y$-direction is about $2$ kilometer per hour which means about $21.6$~seconds to travel the road section from a side to the other side.

The time-dependent microscopic positions and the microscopic velocities of vehicles can be used to compute the macroscopic quantities, namely the \emph{density} (measured as number of vehicles per kilometer), the \emph{flux} (measured as number of vehicles per hour) and the \emph{mean speed} (measured as kilometer per hour) of the flow. We refer to~\cite{HmFaVg} for a detailed description, where we derived the macroscopic data for each direction of the flow, separately.

The diagrams showing the relations between the vehicle density $\rho$ and the fluxes or the mean speeds in the two possible directions are called \emph{fundamental diagrams} and \emph{speed-density diagrams}, respectively. They represent the basic tools for the analysis of traffic problems operating in a homogeneous \emph{steady state} or \emph{equilibrium} conditions.

\begin{figure}[t!]
	\begin{center}
	\includegraphics[width=4.3cm]{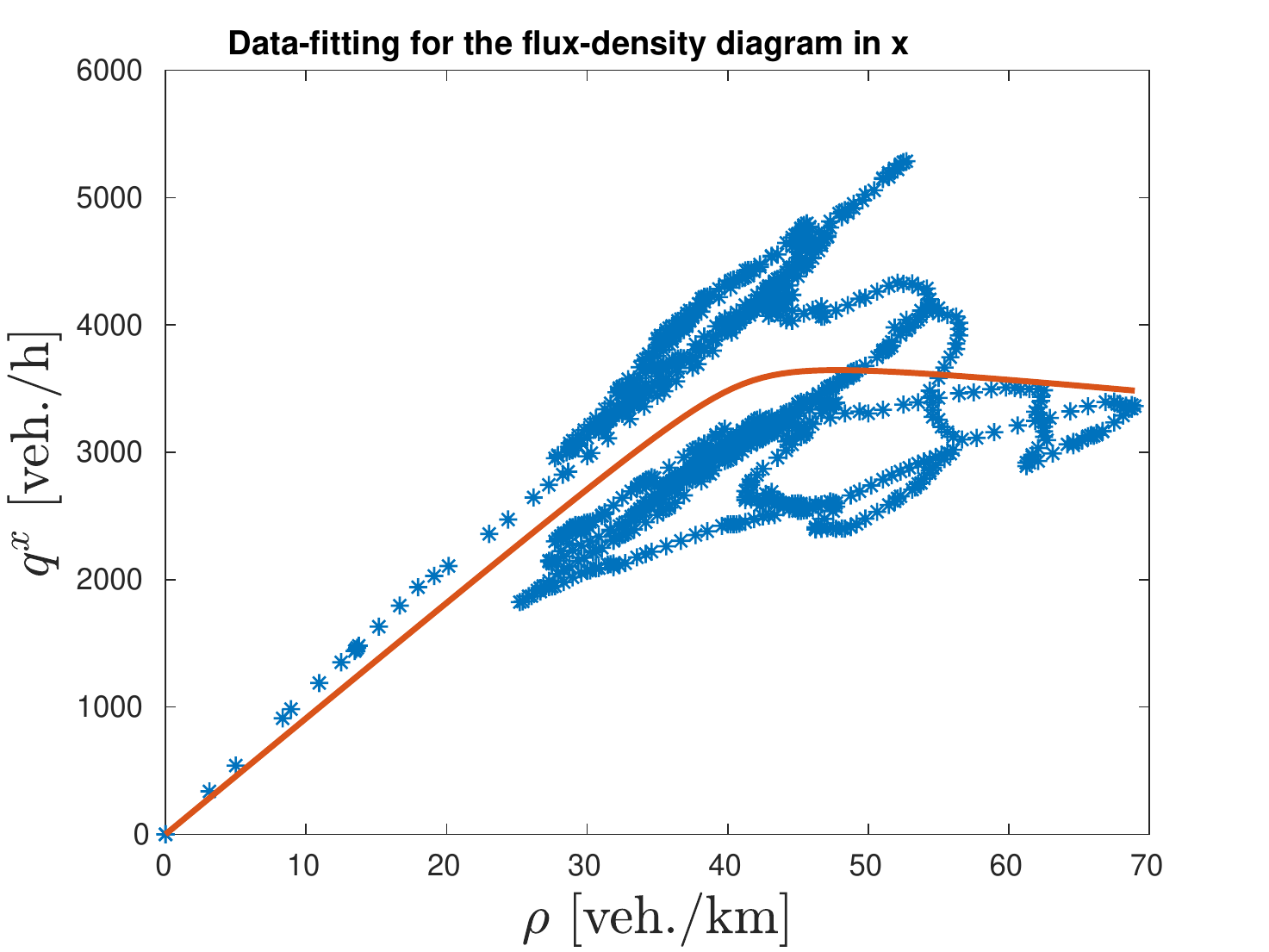}
	\includegraphics[width=4.3cm]{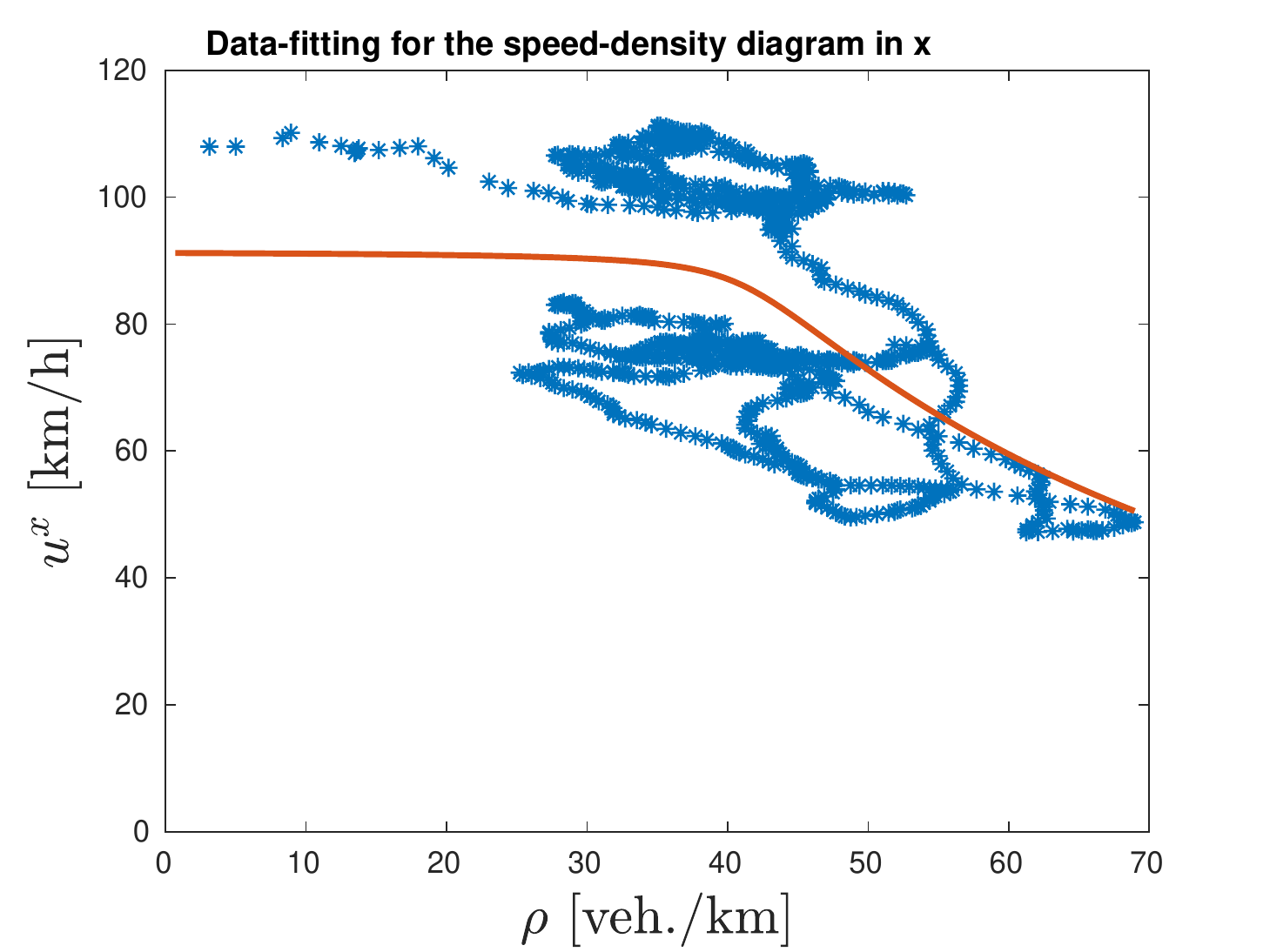}
	\\
	\includegraphics[width=4.3cm]{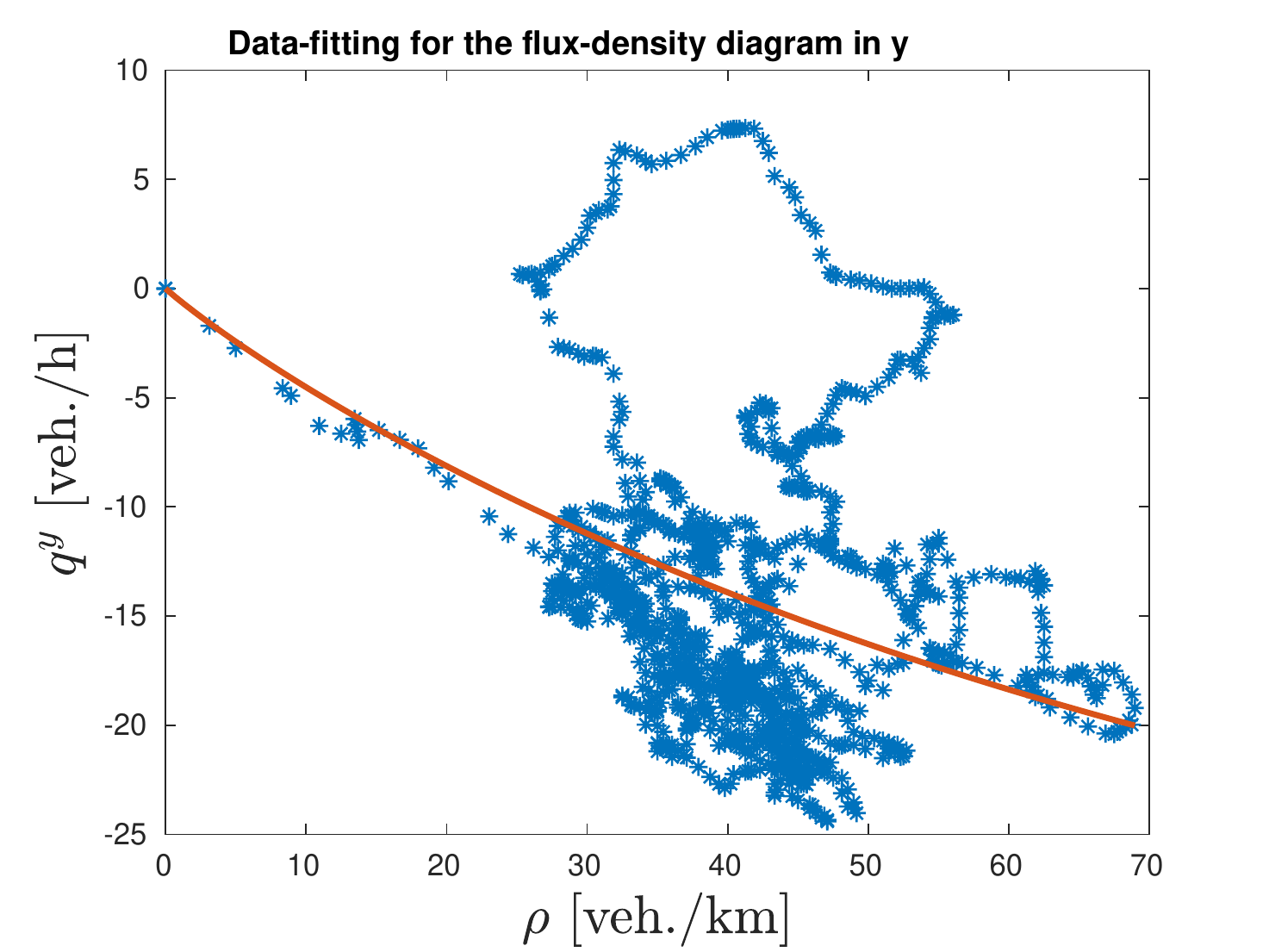}
	\includegraphics[width=4.3cm]{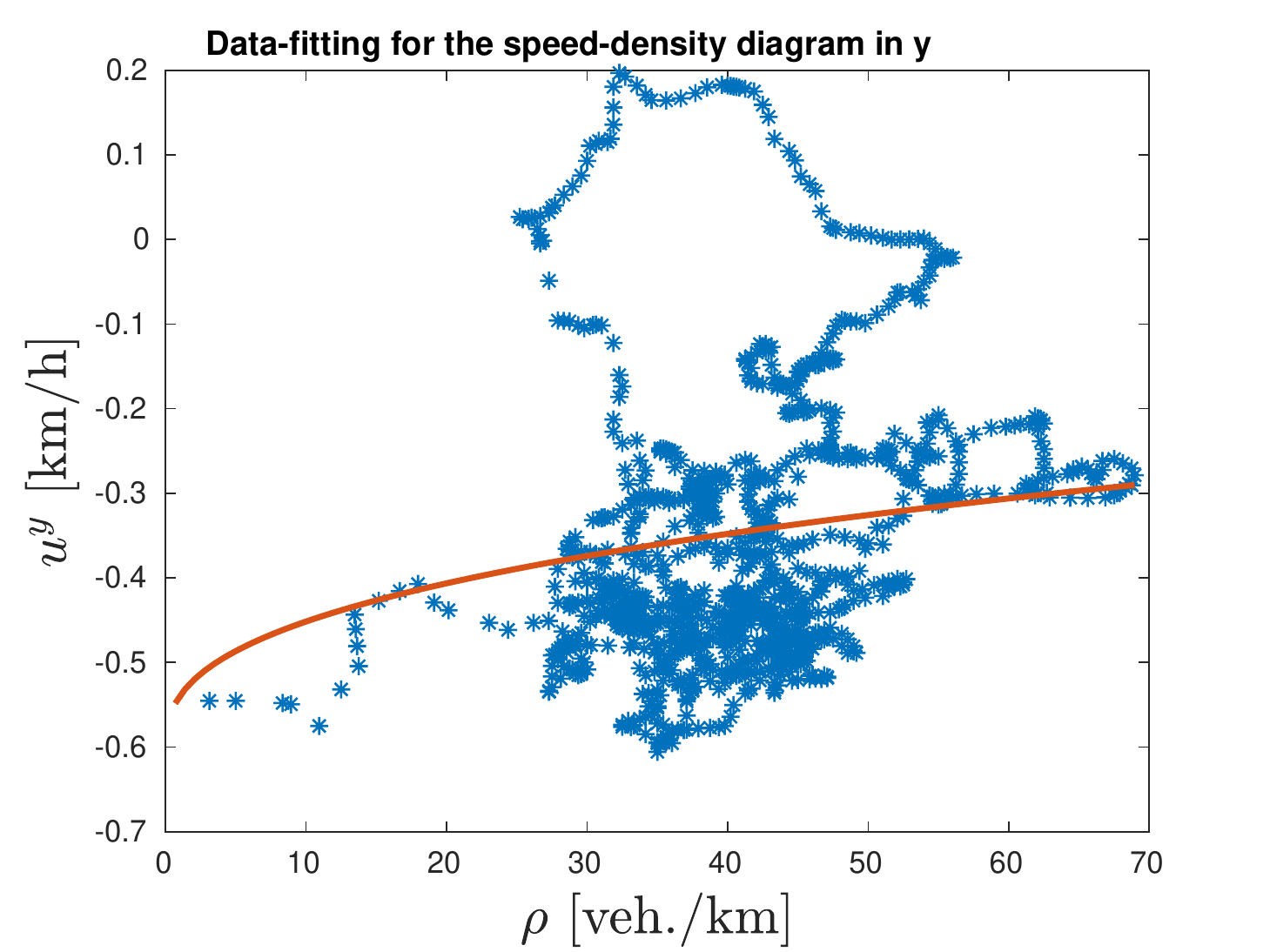}
	\caption{Experimental diagrams from the A3 German highway. Top: flux-density (left) and speed-density (right) diagrams in $x$-direction. Bottom: flux-density (left) and speed-density (right) diagrams in $y$-direction.\label{fig:experimentalFD}}
	\end{center}
\end{figure}

In Figure~\ref{fig:experimentalFD} we show the fundamental diagrams and the speed-density diagrams resulting from the German data-set. The diagrams are obtained by computing the macroscopic quantities each $1$~second and then aggregating the data over a time period of $60$~seconds. The data-set provides several levels of congestion but we never observe bumper-to-bumper conditions. In fact, the maximum density is about $70$~vehicles per kilometer. The red solid lines in each panel of Figure~\ref{fig:experimentalFD} show the best data-fit obtained by solving a constrained minimization problem. See~\cite{HmFaVg} for further details.

\begin{figure}[t!]
	\begin{center}
	\vspace{-0.31cm}
	\includegraphics[width=8.4cm]{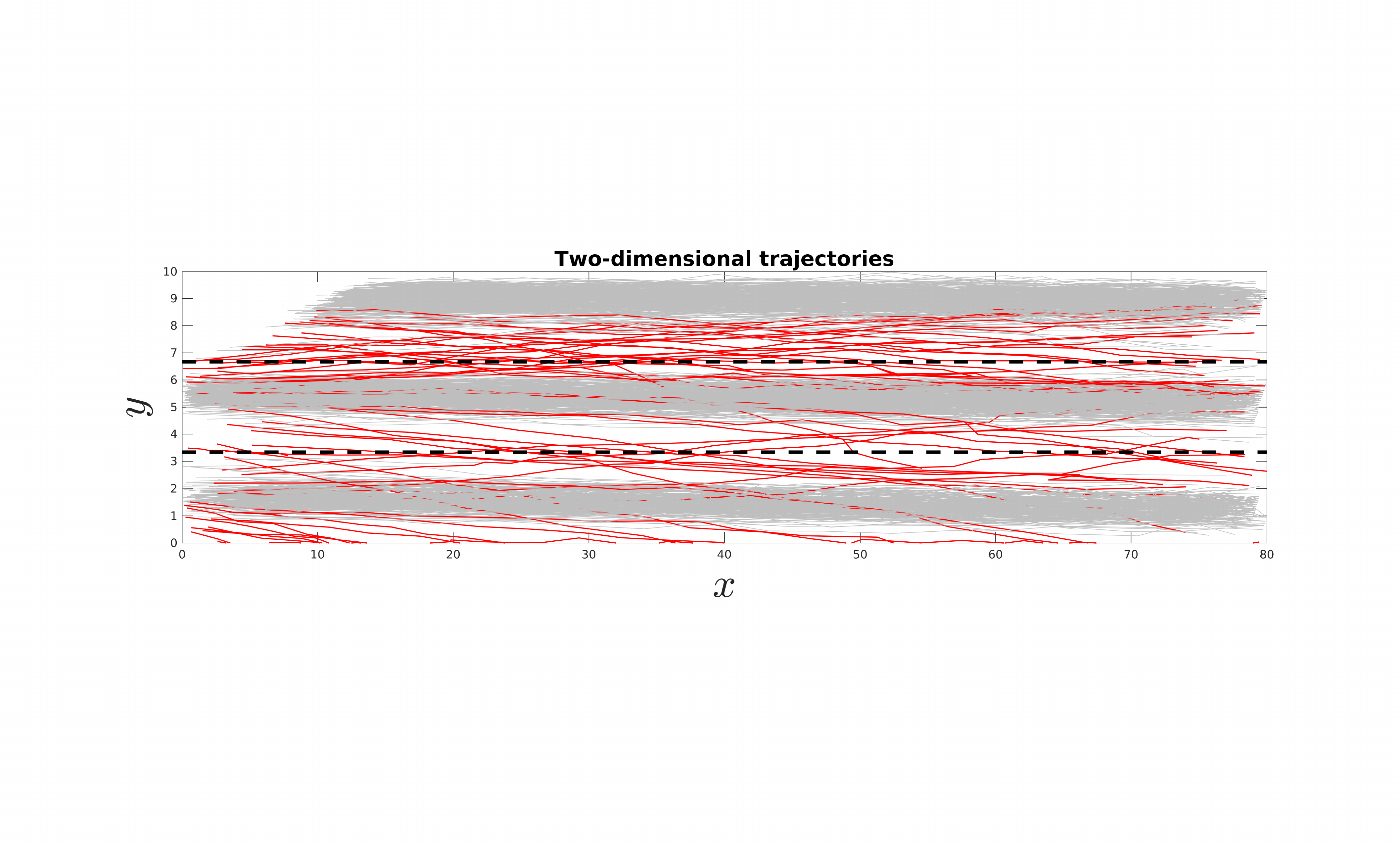}\vspace{-1.5cm}
	\caption{Two-dimensional trajectories extrapolated from the German data-set. In red we show the trajectories of vehicles crossing a lane while traveling.\label{fig:Trajectories}}
	\end{center}
\end{figure}

Notice that the flux and the speed in $y$-direction have positive and negative values since across the lanes vehicles are free to travel in the two directions, towards right and left. Precisely, we assume that positive speeds represent the motion towards the leftmost lane, instead negative speeds represent the motion towards the rightmost lane. The values of the flux and of the mean speed in $y$-direction are about $10^3$ smaller than the values in $x$-direction. This is obvious since the velocity of vehicles along the road is higher then the lateral velocity and thus this latter is not dominant with respect the other speed. In other words, we are looking at two behaviors occurring at different scales. But this does not mean that the behavior in $y$-direction can be neglected and that the flow across the lanes does not influence the safety. In fact, an analysis of the trajectories shows that about the $15\%$ of the total vehicles crosses a lane while traveling the road section. See Figure~\ref{fig:Trajectories}.

\section{Risk analysis on a multi-lane road} \label{sec:TTC}

Several metrics were introduced in order to study risk levels in traffic flow scenarios. Among them, here we will consider the so-called \emph{Time-To-Collision} (TTC) metric. The TTC can be defined as the remaining time until a collision between two vehicles would have occurred, see~\cite{Hayward}. The notion of TTC represents the most used indicator in the analysis of safety measures for traffic.

In the literature, the TTC metric is usually used by considering one-dimensional car-following scenarios and thus it is computed by looking at only the positions along the road, as well as the velocities in $x$-direction, of the two interacting vehicles. Here, instead, we are aimed at studying the risk levels on a multi-lane highway, thus using two-dimensional microscopic data. Therefore, we firstly need to define the concept of a car-following interaction in two-space dimensions and we use the same approach introduced in~\cite{HmMsVg}.

\begin{figure}[t!]
	\begin{center}
	\includegraphics[width=8.4cm]{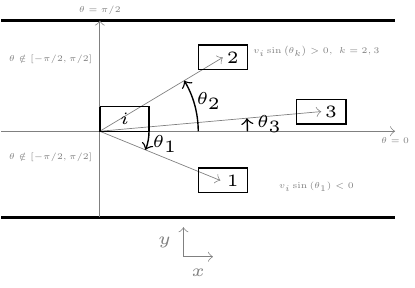}
	\caption{Choice of the interacting car in the case $v_i>0$. The interacting vehicle will be car $2$, namely the nearest vehicle in the driving direction of vehicle $i$.\label{fig:2D:ChoiceOfj}}
	\end{center}
\end{figure}

We assume that there is no particular order among vehicles on the road. We just label them. From now on, let $u_i$ and $v_i$ be the speeds in $x$-direction and $y$-direction, respectively, of car $i$. For each vehicle $i$, the interacting car $j(i)$ is determined by the following map
\begin{equation} \label{eq:FieldCar}
i \mapsto j(i) = \argmin_{\substack{h=1,\dots,N \\ v_i \sin\theta_h > 0 \\ \theta_h\in\left[-\frac{\pi}{2},\frac{\pi}{2}\right]}} \norm{P_h-P_i}_2.
\end{equation}
This choice is motivated as follows, see also Figure~\ref{fig:2D:ChoiceOfj}. Assume that each test vehicle $i$ defines a coordinate system in which the origin is its right rear corner if $v_i\geq 0$ and its left rear corner if $v_i<0$. We are indeed dividing the road in four areas. Let $\theta_h$ be the angle between the $x$-axis (in the car coordinate system) and the position vector $Q_h$ of vehicle $h$. Then the request $\theta_h\in\left[-\frac{\pi}{2},\frac{\pi}{2}\right]$ allows to consider only cars being in front of vehicle $i$. Instead, the request $v_i\sin\theta_h>0$ allows to consider only cars in the driving direction of vehicle $i$. Among all these vehicles we choose the nearest one. Therefore, map~\eqref{eq:FieldCar} can be rewritten as
\begin{equation} \label{eq:FieldCar2}
i \mapsto j(i) = \argmin_{\substack{h=1,\dots,N \\ v_i (y_h - y_i)>0 \\ x_h > x_i}} \norm{P_h-P_i}_2.
\end{equation}

\subsection{Time-To-Collision and Individual Risk} \label{sec:TTCandIR}

Map~\eqref{eq:FieldCar2} allows us to define the car-following scenarios on a two-dimensional road section for each observation time. Here, we are interested in generalizing the TTC metric~\eqref{eq:extendedTTC} for car $i$ in each direction of the flow by looking at the same interacting vehicle $j(i)$. However, since in Section~\ref{sec:dataset} we have described the trajectory of all vehicles with a linear function approximating the time positions in the least-square sense, the acceleration is supposed to be zero. Therefore, we consider a simplified version for the TTC and along with the classical definition of the TTC in $x$-direction
\begin{equation} \label{eq:xTTC}
	\text{TTC}_i^x(t) = 
	 \begin{cases}
	 - \frac{x_{j(i)}-x_i}{u_{j(i)}-u_i}, & \text{if $u_i < u_{j(i)}$}\\[2ex]
	 +\infty, & \text{otherwise},
	 \end{cases}
\end{equation}
in this paper we introduce also the TTC in $y$-direction as
\begin{equation} \label{eq:yTTC}
	\text{TTC}_i^y(t) = 
	\begin{cases}
	- \frac{y_{j(i)}-y_i}{v_{j(i)}-v_i}, & \text{if $\abs{v_i} < \abs{v_{j(i)}}$ or $v_{j(i)}\,v_i<0$}\\[2ex]
	+\infty, & \text{otherwise}.
	\end{cases}
\end{equation}
The $\text{TTC}^y_i$ defines thus the remaining time to a collision with car $j(i)$ moving across the lanes. The analysis of this metric was never considered previously in the literature but we think it is useful to study which direction of the flow mainly influences the safety. Clearly, using map~\eqref{eq:FieldCar2}, a car $i$ could have no leading vehicles, namely when there are no obstacles in its driving direction. Then we set the two TTC metrics to $+\infty$ also in this case.

The TTC metric has been widely used in order to evaluate the risk level in traffic flow phenomena. However, as suggested by several authors, \cite{KuangQuYan}, the TTC does not give direct information on the safety conditions of vehicle. Moreover, the fact that the TTC is $+\infty$ when an interaction does not occur makes this metric not suitable for computations from microscopic data. Therefore, in this paper, following the methodology introduced in~\cite{KuangQuYan}, we represent the \emph{Individual Risk} (IR) described by the TTC by comparing, for each car, this value with thresholds:
\begin{equation} \label{eq:IR}
	\text{IR}_i^{x,y}(t) = \begin{cases}
		\widehat{\text{TTC}}^{x,y} - \text{TTC}_i^{x,y}, & \text{if $\widehat{\text{TTC}}^{x,y} > \text{TTC}_i^{x,y}$}\\[2ex]
		0, & \text{otherwise}.
	\end{cases}
\end{equation}
The fixed values $\widehat{\text{TTC}}^{x,y}$ are the TTC thresholds in the two direction of the flow and they are supposed to be the safety values below which an interaction is considered unsafe. Using definition~\eqref{eq:IR} then a car-following scenario is considered to be safe if the IR is low. Several value of the TTC threshold are proposed, see e.g.~\cite{MinderhoudBovy,KuangKuWang,KuangEtAl}. Here, we consider safety values the times to travel the road section in $x$ and $y$ at the maximum speeds detected in the two directions. Thus, using the German data-set introduced in Section~\ref{sec:dataset} we have
$$
	\widehat{\text{TTC}}^x = 2.7~\text{seconds}, \quad \widehat{\text{TTC}}^y = 21.6~\text{seconds}.
$$
With this choice we say that a vehicle is in safe conditions if it can freely travel along and across the road at the maximum speed.

In order to aggregate the individual risk of single car-following scenarios we proceed as follows. First we fix a sequence of $M+1$ equally spaced discrete times $\{t_k\}_{k=0}^M$ such that $t_{k+1}-t_k=dt$, $t_0=0$ and $t_M=t_{\max}$, where $t_{\max}$ is the final observation time in the data-set (here 20 minutes). Then, for each discrete time $t_k$, we count the number of vehicles $N(t_k)$ and for each one we find the interacting vehicle using~\eqref{eq:FieldCar2}. We define the individual risk at time $t_k$ as
\[
	\widetilde{\text{IR}}^{x,y}(t_k) = \frac{1}{N(t_k)}\sum_{i=1}^{N(t_k)} \text{IR}_i^{x,y}(t_k), \quad k=0,\dots,M
\]
where $\text{IR}_i^{x,y}$ is given by~\eqref{eq:IR}. Finally, we consider a moving mean by aggregating with respect a certain time period $T$, with $T \ll t_{\max}$ and including $m$ consecutive observations. This temporal average leads to $\left\lceil \frac{M+1}{m} \right\rceil+1$ values of the individual risk
\begin{equation} \label{eq:aggregatedIR}
	\text{IR}^{x,y}_{k_0} = \frac{1}{T} \sum_{k=k_0}^{k_0+m-1} \widetilde{\text{IR}}^{x,y}(t_k), \quad k_0 = 0,\dots,\left\lceil \frac{M+1}{m} \right\rceil.
\end{equation}
In particular, in this paper we take $dt=1$ second and then we aggregate the data over the time period $T=30$ or $T=60$ seconds.

\subsection{Macroscopic formulation of TTC} \label{sec:macro}

Here we show, from a mathematical point-of-view, that the TTC is indeed able to describe the safety levels of a road since it is linked to the variation of the traffic volume. For the sake of simplicity, let us to focus on the simple one-dimensional case and thus we assume to have $\text{TTC}_i^y = 0$, $\forall i$. We define the density and the specific volume around each vehicle at time $t$ as
\[
	\rho_i(t) = \frac{\Delta X}{x_{j(i)}(t)-x_i(t)}, \quad \tau_i(t) = \frac{1}{\rho_i(t)},
\]
where $\Delta X$ is the characteristic length of a vehicle. Differentiating $\tau_i$ in time, we have that the variation of the specific volume is given by
\[
	\dot{\tau}_i(t) = \frac{u_{j(i)}-u_i}{\Delta X}.
\]
Let us again to consider the simplified version of the TTC~\eqref{eq:xTTC}, namely we assume that the velocity of each car on the road is constant while traveling the road section. Using the same considerations introduced in~\cite{HmMsVg} in order to derive macroscopic models from particle models, we can identify the microscopic velocities $u_i$ with a function $u$ out of the discrete dynamics in such a way $u_i = u(x_i)$, $\forall i$. Moreover, at the same time we can identify $\tau_i$ with a function $\tau$ out of the discrete dynamics in such a way $\tau_i = \tau(x_i,t)$. With these considerations we can write
\begin{align*}
	\frac{\mathrm{d}}{\mathrm{d}t} \tau(x_i,t) &= \frac{u(x_{j(i)})-u(x_i)}{\Delta X} \\ \text{TTC}_i & = - \frac{x_{j(i)}-x_i}{u(x_{j(i)})-u(x_i)}.
\end{align*}
Since we are interested in the macroscopic formulation, i.e. when the number of vehicles becomes larger, it is quite natural to assume that $\Delta X \approx x_{j(i)}-x_i$. In the macroscopic limit we consider the limit for $\Delta X$ going to zero and thus we obtain
\[
	\frac{\mathrm{d}}{\mathrm{d}t} \tau(x,t) = \partial_x u(x) \quad \text{and} \quad \text{TTC} \approx - \frac{1}{\partial_x u(x)}.
\]
In other words, the inverse of the TTC is able to describe the time evolution of the specific volume and therefore this in turn means that there is a strict link between the safety and the congestion level of the road. This result can be easily extended also to the two-dimensional case, using the considerations introduced in~\cite{HmMsVg}. In addition, the above result paves the way to a possible real-time analysis of the potential conflicts on a road by means of macroscopic models.

\section{Potential risk on the German highway} \label{sec:simulations}

In this section we show that, at least using the German data-set, the safety on a multi-lane highway seems to be highly influenced by the movements in the orthogonal direction of the flow, i.e. by the lane changing.

\begin{figure}[t!]
	\begin{center}
		\includegraphics[width=4.3cm]{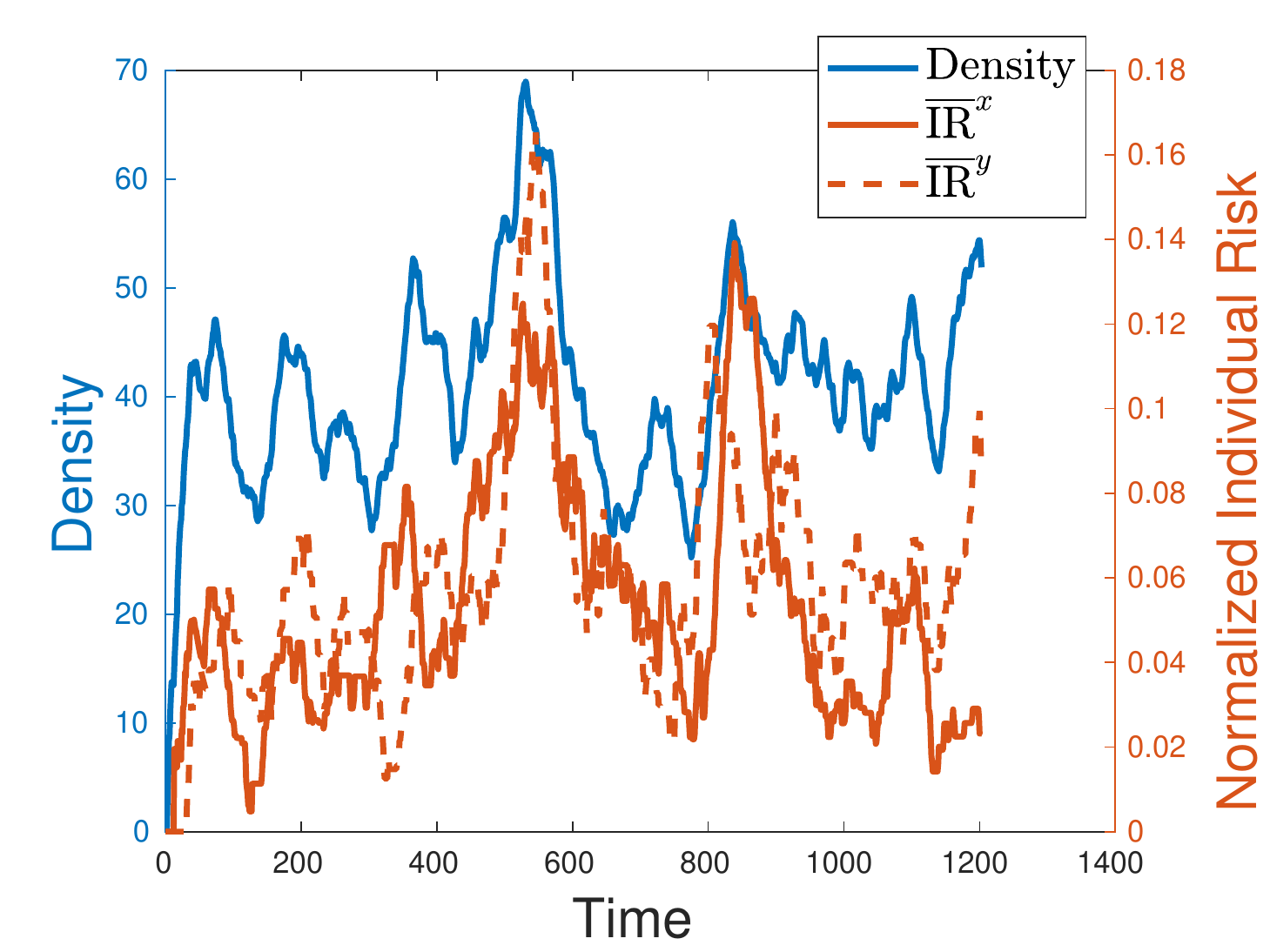}
		\includegraphics[width=4.3cm]{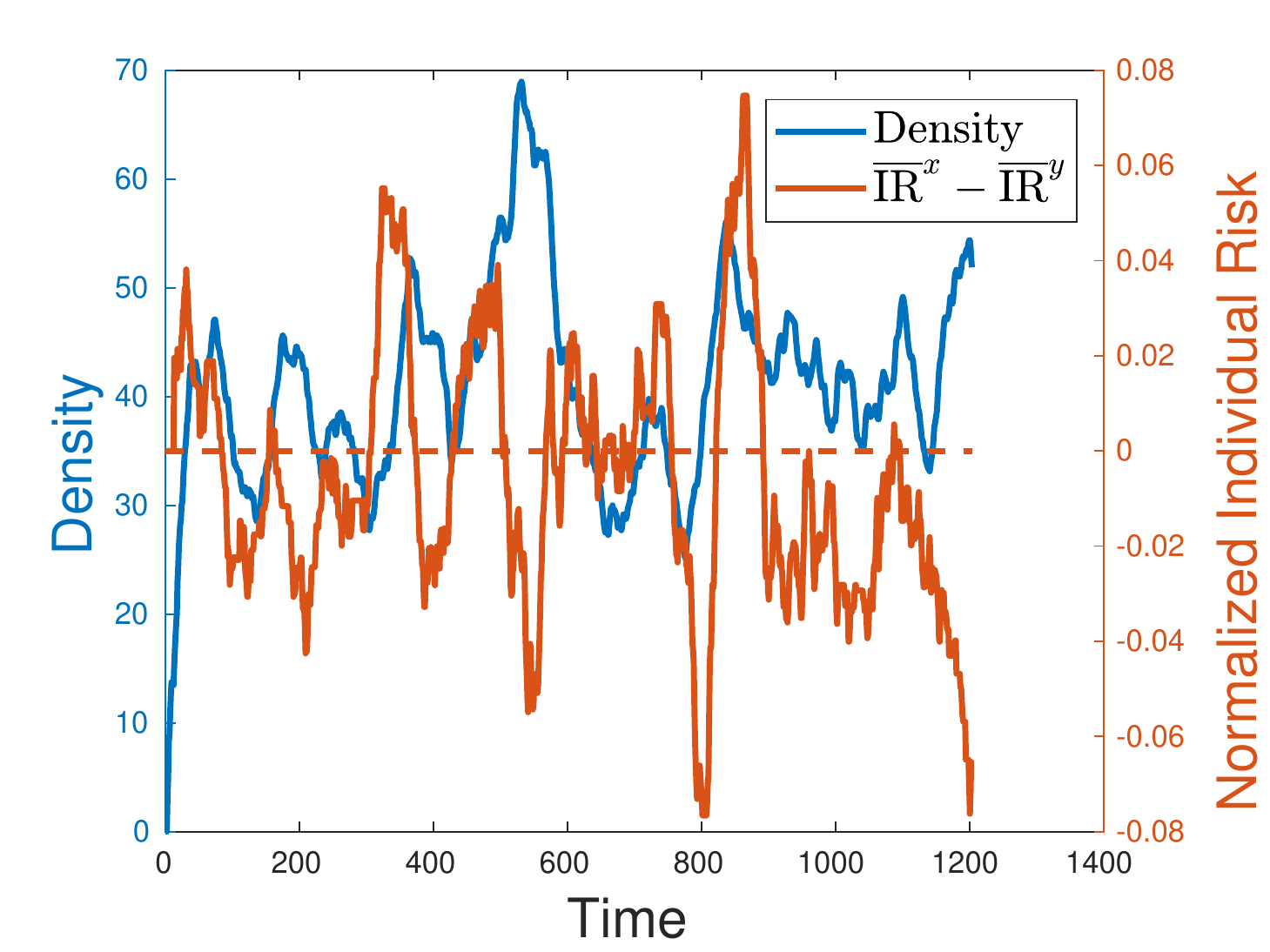}
		\caption{Left: variation in time of the density (blue line) and of the normalized individual risks $\overline{\text{IR}}^x$ (solid red line) and $\overline{\text{IR}}^y$ (dashed red line). Right: variation in time of the density (blue line) and of the difference $\overline{\text{IR}}^x-\overline{\text{IR}}^y$ of the normalized individual risks (solid red line).\label{fig:timeIR}}
	\end{center}
\end{figure}

In Figure~\ref{fig:timeIR} we consider the time behavior of the aggregated individual risk~\eqref{eq:aggregatedIR} for both directions. However, in order to compare the two quantities we show the normalized values with respect to the safety thresholds, namely 
\begin{equation} \label{eq:normIR}
	\overline{\text{IR}}^{x,y}_{k_0} = \frac{\text{IR}^{x,y}_{k_0}}{\widehat{\text{TTC}}^{x,y}} \in [0,1].
\end{equation}
In this way the individual risk is directly comparable and quantities~\eqref{eq:normIR} give, in probability, the risk level on the road. In the left plot of Figure~\ref{fig:timeIR} we show the time variation of the normalized individual risks $\overline{\text{IR}}^x$ (solid red line) and $\overline{\text{IR}}^y$ (dashed red line). The values of the individual risk are given on the right $y$-axes. The blue solid line, instead, shows the time variation of the density and the values are given on the left $y$-axes. It is clear that both individual risk increase for higher values of the density. In the right plot of Figure~\ref{fig:timeIR}, along with the density, we show the difference $\overline{\text{IR}}^x-\overline{\text{IR}}^y$ of the normalized individual risks. Thus, values below zero (see the dashed red line) mean that the individual risk in $y$-direction is higher than the individual risk in $x$-direction. Using this plot we can easily observe that the normalized individual risk $\overline{\text{IR}}^y$ is higher in connection to the peaks of the density. This behavior can be explained by the fact that more lane changing occur when the density increases.

The above consideration can be also quantitatively analyzed by using a methodology used e.g. recently in~\cite{KuangQuYan} and which consists in dividing the aggregated values of the individual risk into many traffic states, sorted by the density, with uniform span. With this approach it is possible to analyze easily the relationship between traffic risk and traffic states. The following procedure allows us to divide the aggregated macroscopic data into many traffic states, sorted by the density.
\begin{description}
	\item[Step 1] Rank all the $\left\lceil \frac{M+1}{m} \right\rceil+1$ observations of the individual risks according to the related value of the density, from the lowest to the highest.
	\item[Step 2] Count the total number $\widetilde{n}$ of intervals for those density observations with a constant span $\delta$. Thus, 
	\[
		\widetilde{n} = \text{round}\left(\frac{\rho_{\max}-\rho_{\min}}{\delta}\right)
	\]
	where $\rho_{\max}$ and $\rho_{\min}$ are the maximum and the minimum value of the density computed from the data-set.
	\item[Step 3] Find the range of the intervals as
	\[
		[\rho_{\min}+(n-1)\delta,\rho_{\min}+n\delta], \quad n=1,\dots,\widetilde{n}
	\]
	and then count the number $N_n$, $n=1,\dots,\widetilde{n}$, of density data belonging to each interval.
	\item[Step 4] Using the aggregated individual risk defined in~\eqref{eq:aggregatedIR}, compute the Cumulative Risk (CR) for each interval as
	\begin{equation} \label{eq:CR}
		\text{CR}^{x,y}_n = \sum_{i=K_{n-1}}^{K_{n-1}+N_n} \text{IR}^{x,y}_i, \; K_{n-1} = \sum_{k=1}^{n-1} N_k, \; n=1,\dots,\widetilde{n}
	\end{equation}
	where, obviously, we have
	\[
		\sum_{k=1}^{\widetilde{n}} N_k = \left\lceil \frac{M+1}{m} \right\rceil+1.
	\]
	\item[Step 5] Finally, compute the Average Risk (AR) value for each interval as
	\begin{equation} \label{eq:AR}
		\text{AR}^{x,y}_n = \frac{\text{CR}^{x,y}_n}{N_n}, \quad n=1,\dots,\widetilde{n}.
	\end{equation}
\end{description}
The same procedure described above can be used to sort the traffic states with respect to the other macroscopic fundamental quantities, as the speeds and the fluxes. In the following simulations, in place of the average risk $\text{AR}^{x,y}$ we will consider the normalized average risk defined as
\begin{equation} \label{eq:normalizedAR}
	\overline{\text{AR}}^{x,y}_n = \frac{\text{AR}^{x,y}_n}{\widehat{\text{TTC}}^{x,y}}, \quad n=1,\dots,\widetilde{n}
\end{equation}
in order to make comparable the quantities in the two directions of the flow.

\begin{figure}[t!]
	\begin{center}
		\includegraphics[width=4.3cm]{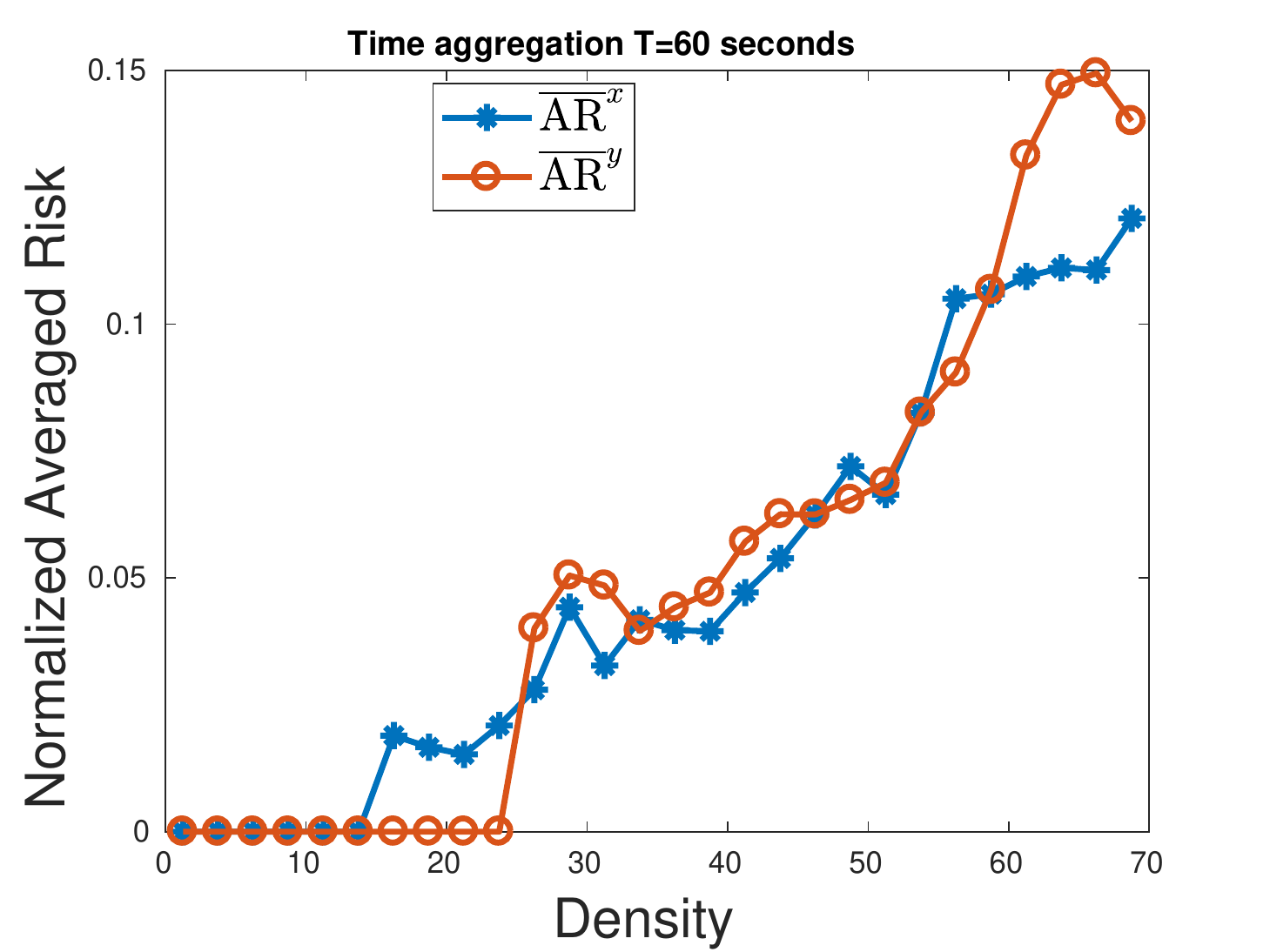}
		\includegraphics[width=4.3cm]{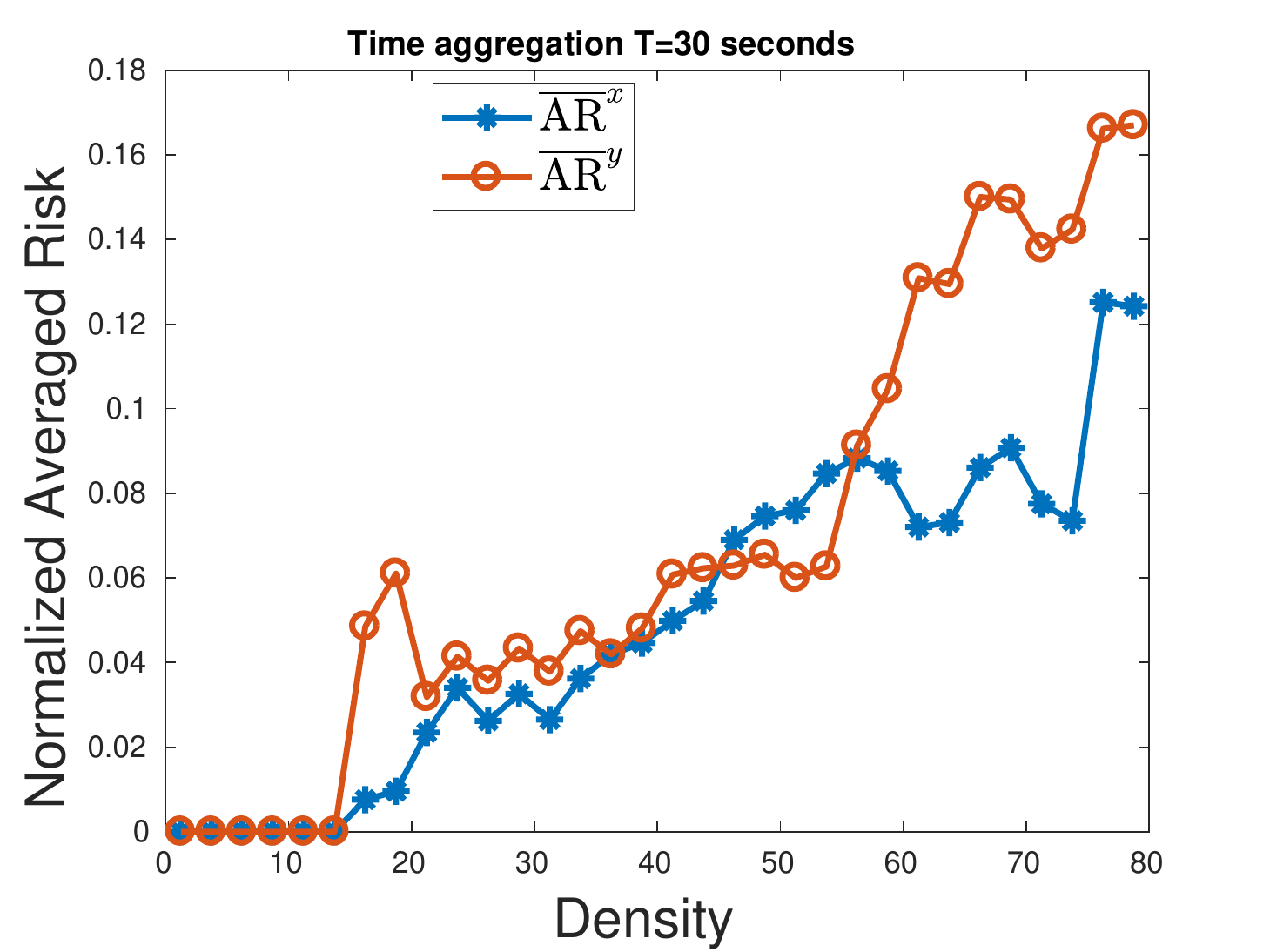}
		\caption{The normalized average risk in the $x$-direction (blue *-symbols) and in the $y$-direction (red circles) of the flow computed on the traffic states sorted by the density. Left: time aggregation $T=60$~seconds. Right: time aggregation $T=30$~seconds.\label{fig:rhoAR}}
	\end{center}
\end{figure}

\begin{figure}[t!]
	\begin{center}
		\includegraphics[width=4.3cm]{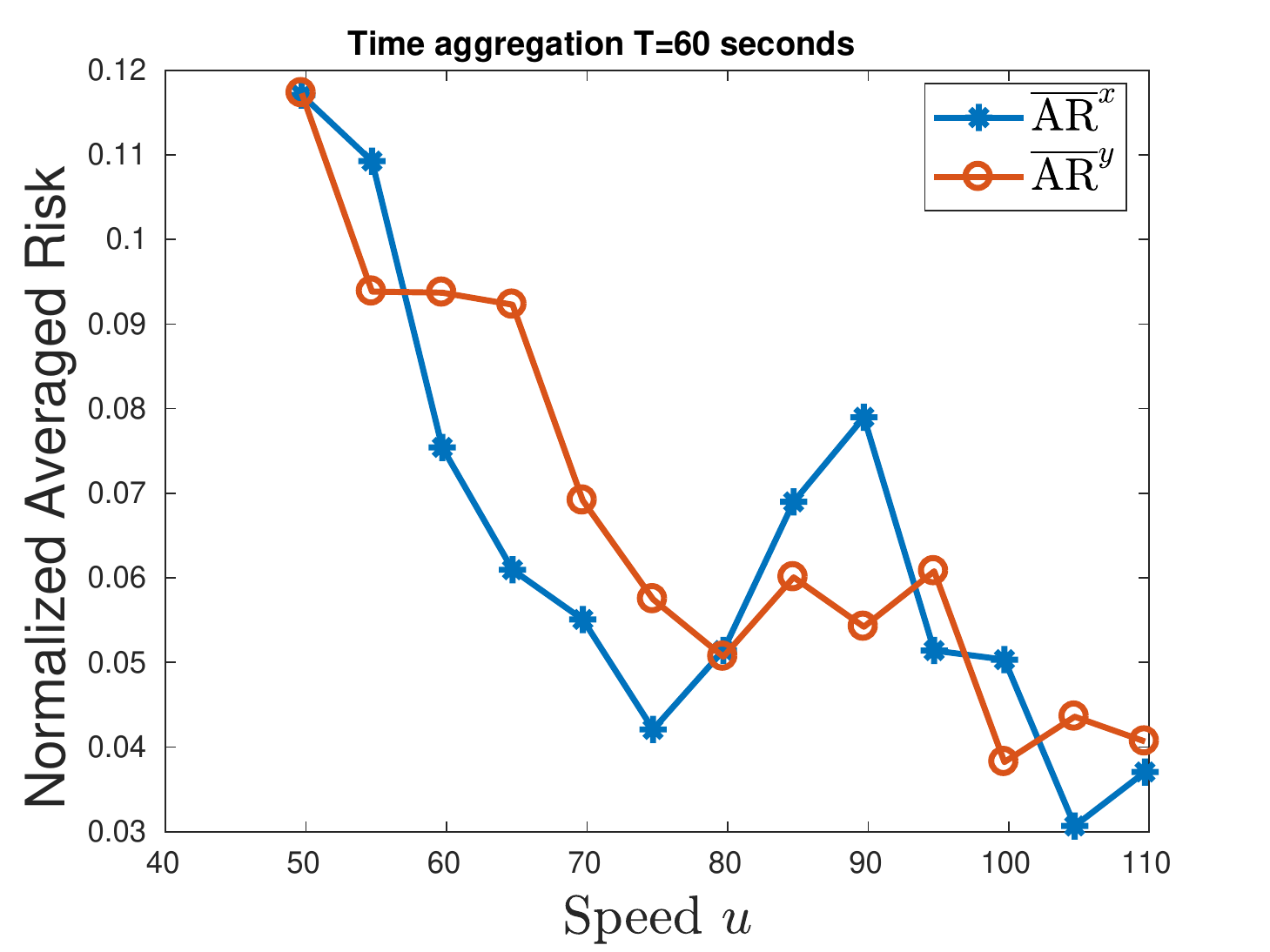}
		\includegraphics[width=4.3cm]{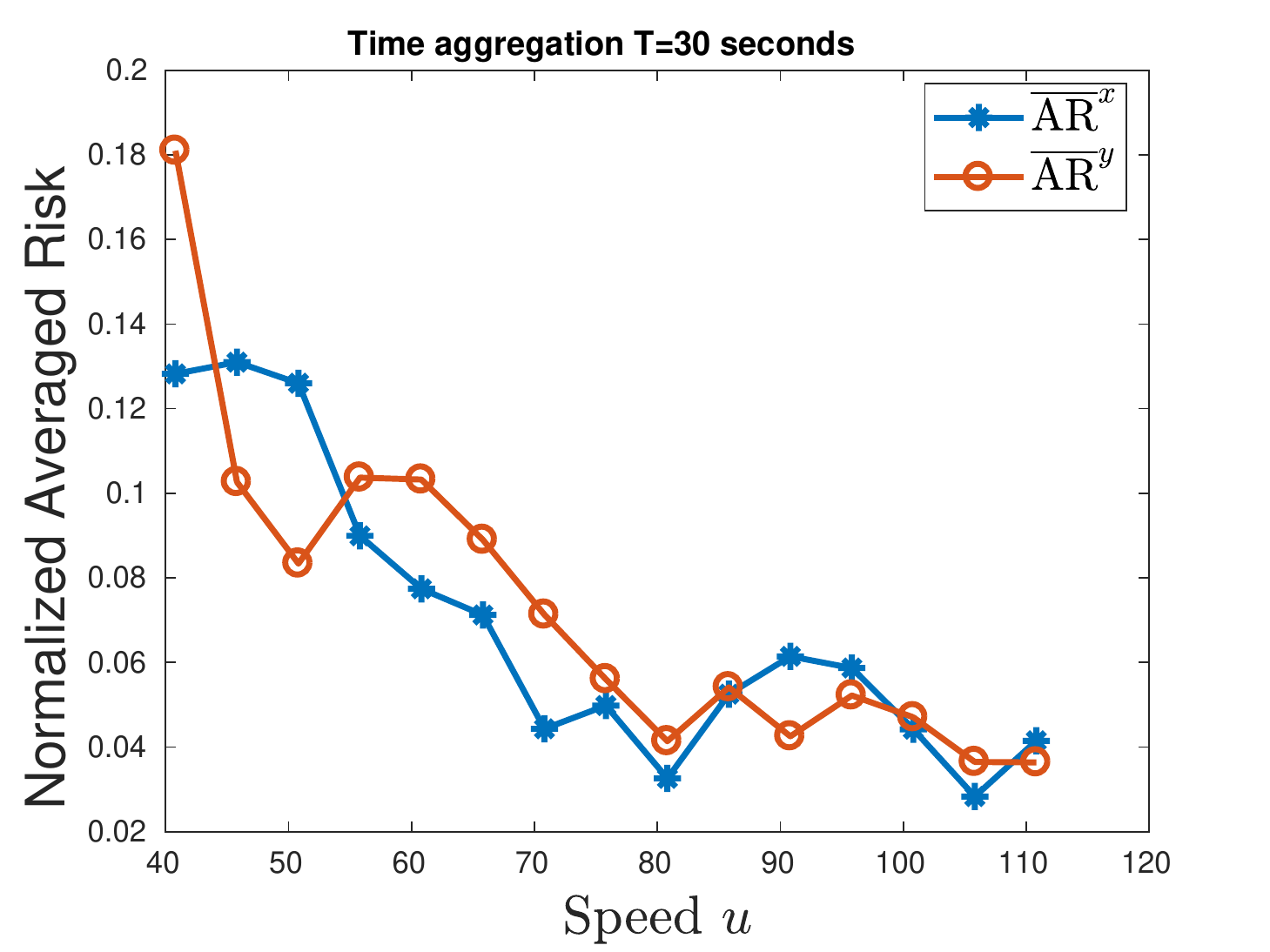}
		\caption{The normalized average risk in the $x$-direction (blue *-symbols) and in the $y$-direction (red circles) of the flow computed on the traffic states sorted by the speed $u$ along the road. Left: time aggregation $T=60$~seconds. Right: time aggregation $T=30$~seconds.\label{fig:uAR}}
	\end{center}
\end{figure}

\begin{figure}[t!]
	\begin{center}
		\includegraphics[width=4.3cm]{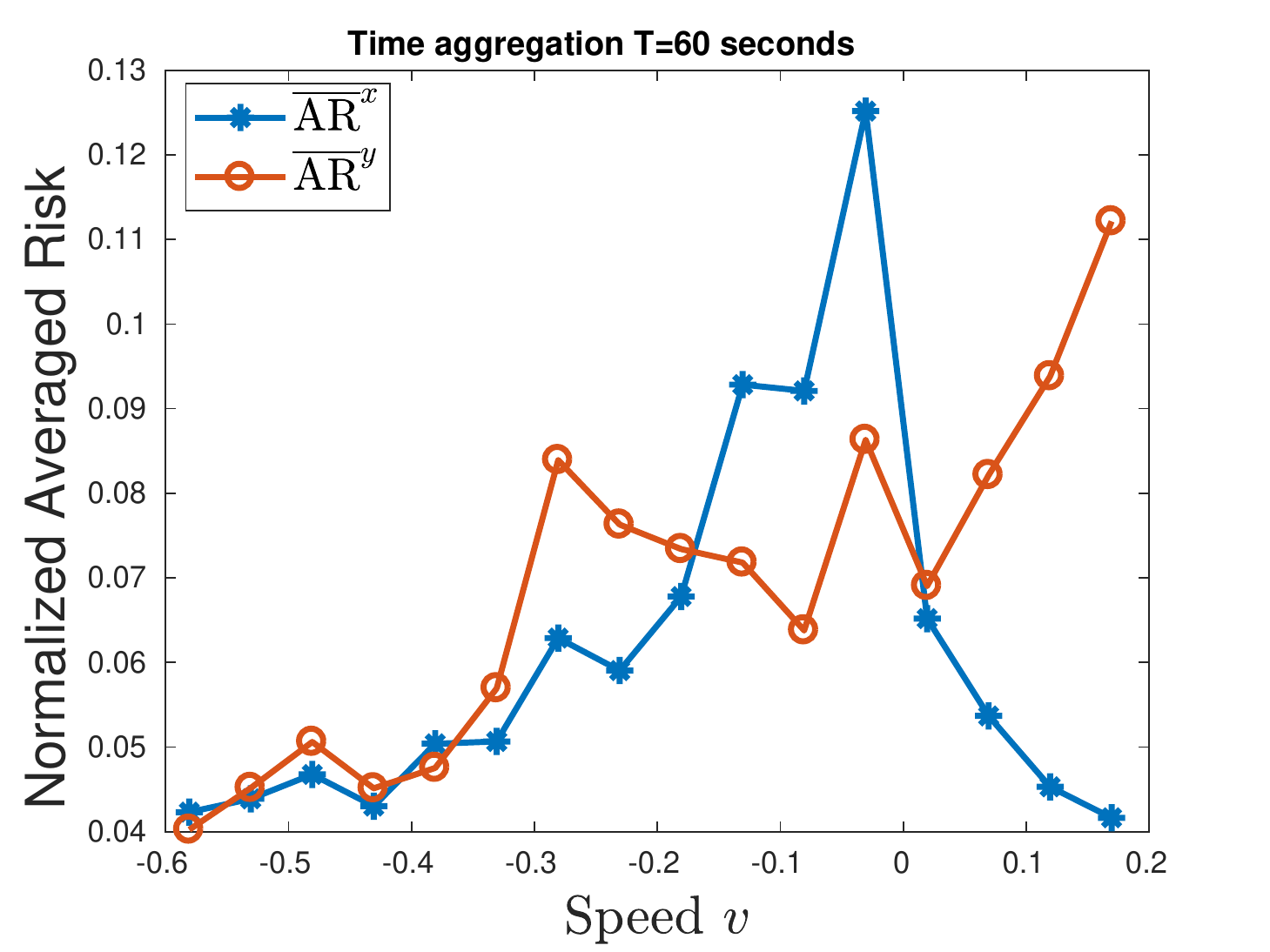}
		\includegraphics[width=4.3cm]{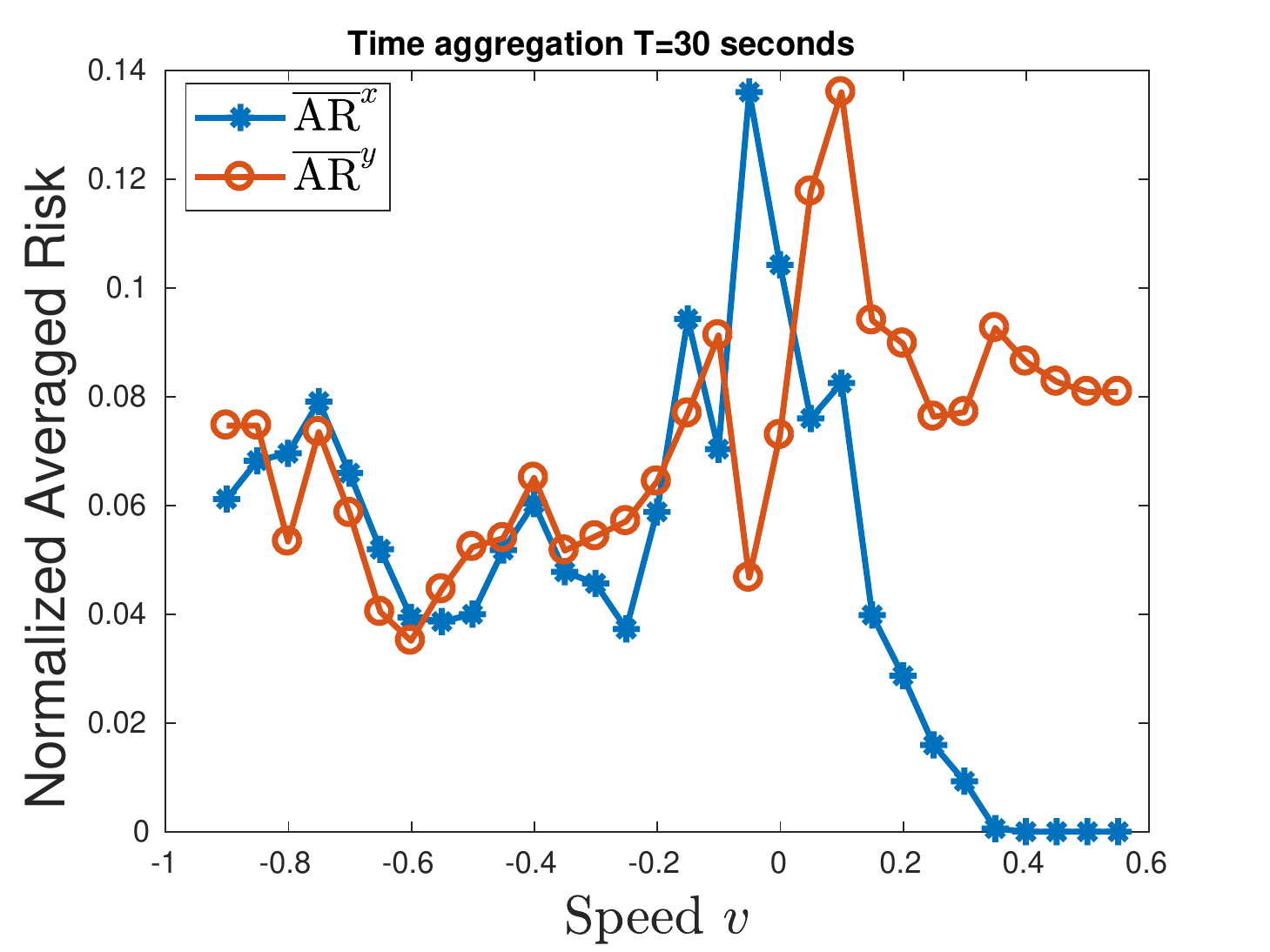}
		\caption{The normalized average risk in the $x$-direction (blue *-symbols) and in the $y$-direction (red circles) of the flow computed on the traffic states sorted by the speed $v$ across the lanes. Left: time aggregation $T=60$~seconds. Right: time aggregation $T=30$~seconds.\label{fig:vAR}}
	\end{center}
\end{figure}

\begin{figure}[t!]
	\begin{center}
		\includegraphics[width=4.3cm]{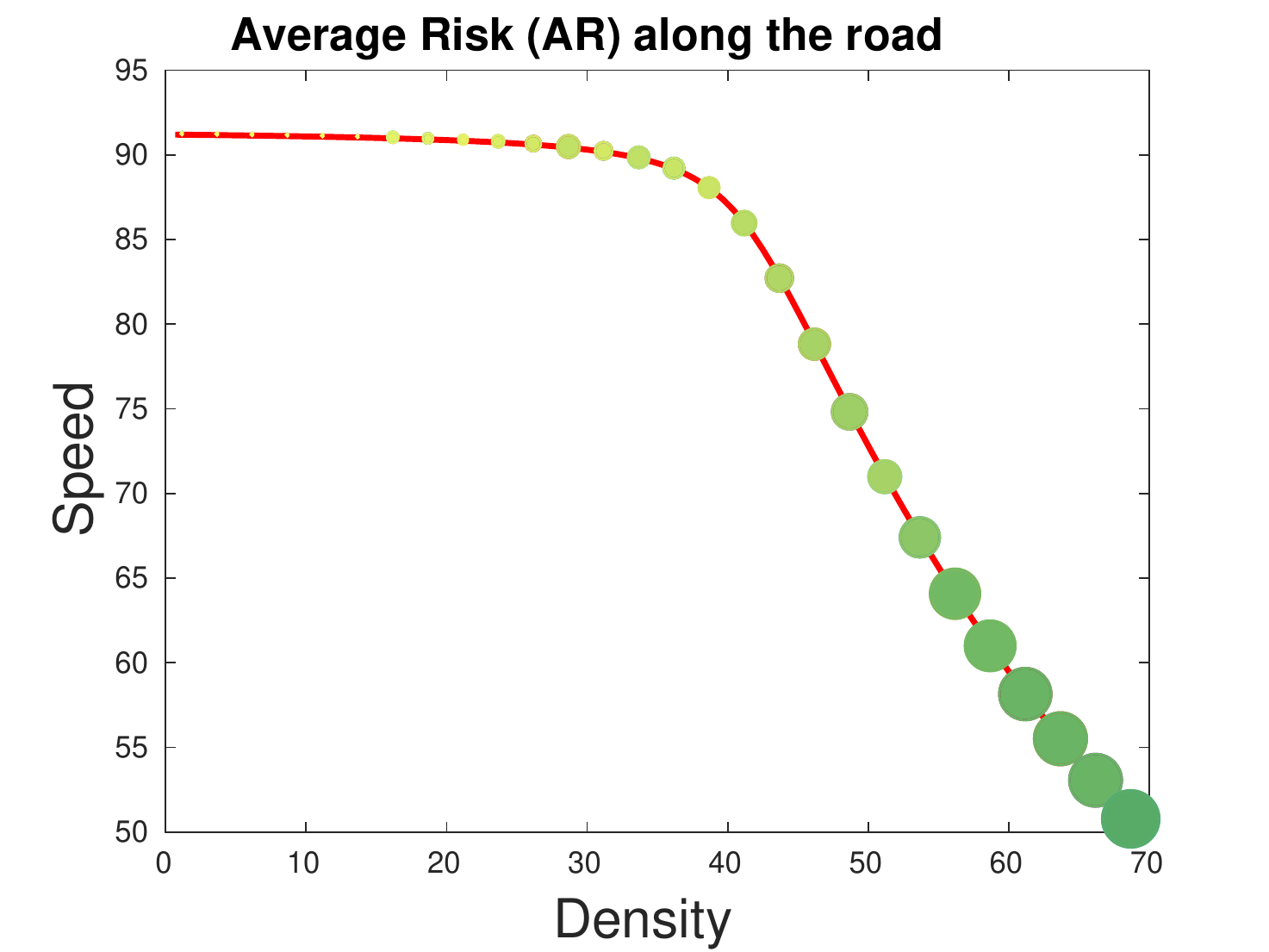}
		\includegraphics[width=4.3cm]{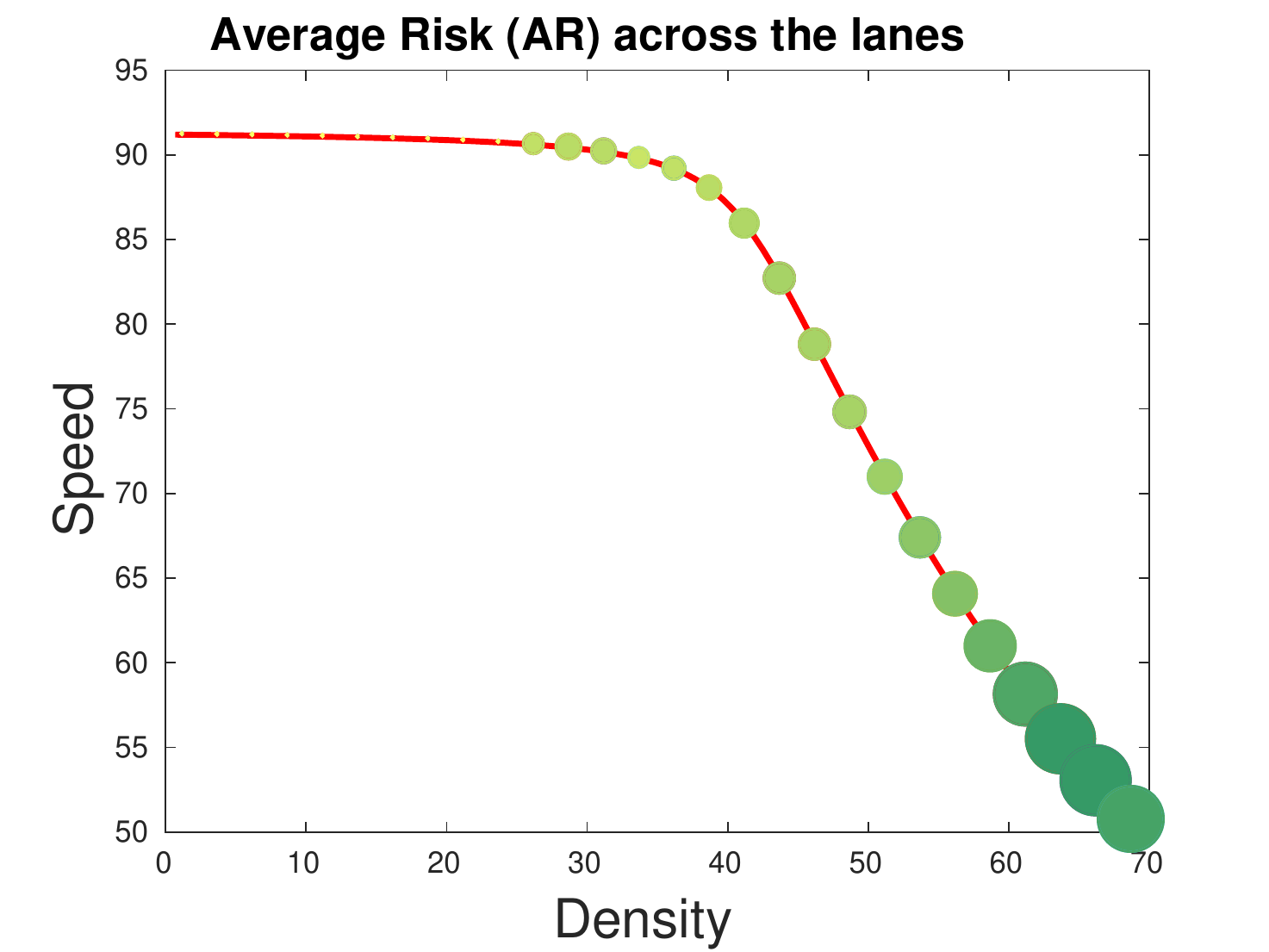}
		\caption{The normalized average risk in the $x$-direction (left) and in the $y$-direction (right) computed on the traffic states sorted by the density and mapped onto the best fitting of the speed-density diagram in $x$-direction.\label{fig:ARonFD}}
	\end{center}
\end{figure}

In Figure~\ref{fig:rhoAR} we show the normalized average risk $\overline{\text{AR}}^x$ and $\overline{\text{AR}}^y$ computed on the traffic states sorted by the density as described by the previous procedure. In particular, the traffic states are obtained by using $\delta = 2.5$ vehicles per kilometer, which represents the minimum value of span to get at least one observation for each interval defined in Step 3. The left and the right panels of Figure~\ref{fig:rhoAR} differ for the time aggregation, which is taken as $T=60$ seconds and $T=30$ seconds, respectively. We observe that the result does not change by varying $T$. In fact, in both panels we observe that the average risk increases as the traffic density increases. This means that a higher density influences the safety and leads to more conflicts. Moreover, the two average risks are comparable when the density is still low, while for higher values $\overline{\text{AR}}^y$ becomes larger than $\overline{\text{AR}}^x$ proving that the traffic conflicts are mainly due to the interactions in $y$-direction, thus regarding lane changing.

The above consideration is explained also by Figure~\ref{fig:uAR} and Figure~\ref{fig:vAR} in which we show the normalized average risk $\overline{\text{AR}}^x$ and $\overline{\text{AR}}^y$ computed on the traffic states sorted by the two speed in the two directions of the flow. We take here $\delta$ as $7.5$ and $0.05$ kilometer per hour, respectively. We observe that the risks decreases as the speed along the road increases, i.e. when the density is low and thus vehicles are free to travel with higher speeds. More significant is~Figure~\ref{fig:vAR}. In fact, we notice that the risk in $x$- and $y$-direction is low for negative values of the lateral speed. This happens when the density is low and vehicles tend to reach the slowest lane. In contrast, the risk in $\overline{\text{AR}}^x$ increases and is higher than $\overline{\text{AR}}^y$ around the zero values of the lateral speed, i.e. when the lane changing are rare and the conflicts mainly happen with cars ahead. On the opposite side, when the lateral speed becomes positive, i.e. when the higher values of the density induce vehicles to overtake and to move from the slowest lane, then the average risk in $y$-direction increases substantially. This behavior again proves that the flow in $y$-direction affects the traffic safety more than the interactions occurring along the road.

Finally, in Figure~\ref{fig:ARonFD} we again show the average risk computed on the traffic states sorted by the density but for each state we map the AR value to the corresponding position on the speed-density curve in $x$-direction obtained by fitting the experimental data. In the left panel of Figure~\ref{fig:ARonFD} we show $\overline{\text{AR}}^x$, while in the right one we consider $\overline{\text{AR}}^y$. We use the AR value to determine the size and color of each point. The bigger size and darker color indicate the higher
average risk of a traffic state. Obviously, as observed in Figure~\ref{fig:rhoAR} and Figure~\ref{fig:uAR}, the traffic conflict risk increases with an increase in density and a decrease in speed. Moreover, we notice that the average risk tends to increase substantially when the so-called critical density, namely the density value in which we have a sharp decrease in the speed, is reached.

\section{Outlook} \label{sec:outlook}

In this work we have proposed an analysis of the risk on a multi-lane highway by means of a suitable indicator and studying separately the potential conflicts that could arise from the two direction of the flow. To this end, we have used the Time-To-Collision indicator, which represent the remaining time to avoid a collision between two vehicles. The aim of this paper was to investigate the influence of the lane changing on the traffic safety since only the behavior of the flow along the road was previously considered in the literature. The analysis is carried out using a data-set collected in Germany on the A3 highway. The microscopic trajectories shows that the lane changing is an important behavior in traffic flow and thus the safety can be highly influenced by the movement in the orthogonal direction of the motion of vehicles. This conjecture has been indeed proved by comparing the Time-To-Collision indicators in the two flow directions, showing that the potential conflicts arising by the lane changing are much higher.

In future works, more data can be collected or a different data-set can be considered. Moreover, the analysis conducted here shows the necessity of establishing suitable safety measures having a real impact on the potential risk on multi-lane highways where the flow across the lanes is high. To this end the macroscopic formulation of the Time-To-Collision indicator proposed in this paper could help to develop real-time models to manage safety measures in traffic flow. Ideally, one would measure online traffic data, predict the next minutes of traffic flow and indicate with the macroscopic formulation of risk metrics the likelihood of an accident. Then, by variable speed control or other measures one would try to regulate the traffic flow such that the the Time-To-Collision metric is reduced under the same simulation. Another possibility is the planning process in which we would simulate (e.g. by means of a simulator) a real traffic road and then show the risk indicators. Hopefully, by clever designs of the road the potential conflicts can be reduced. This perspective could be used to plan e.g. construction sites more efficiently.

%\subsection{References}
%
%Use Harvard style references (see at the end of this document). With
%\LaTeX, you can process an external bibliography database 
%using \texttt{bibtex},\footnote{In this case you will also need the \texttt{ifacconf.bst}
%file, which is part of the \texttt{ifaconf} package.}
%or insert it directly into the reference section. Footnotes should be avoided as
%far as possible.  Please note that the references at the end of this
%document are in the preferred referencing style. Papers that have not
%been published should be cited as ``unpublished''.  Capitalize only the
%first word in a paper title, except for proper nouns and element
%symbols.

\begin{ack}
We thank the ISAC institute at RWTH Aachen, Prof. M.~Oeser, MSc. A.~Fazekas and MSc. F.~Hennecke for kindly providing the trajectory data.
\end{ack}

\bibliography{ifacconf,completeBibTex,references}            % bib file to produce the bibliography
                                                     % with bibtex (preferred)
                                                   
%\begin{thebibliography}{xx}  % you can also add the bibliography by hand

%\bibitem[Able(1956)]{Abl:56}
%B.C. Able.
%\newblock Nucleic acid content of microscope.
%\newblock \emph{Nature}, 135:\penalty0 7--9, 1956.

%\bibitem[Able et~al.(1954)Able, Tagg, and Rush]{AbTaRu:54}
%B.C. Able, R.A. Tagg, and M.~Rush.
%\newblock Enzyme-catalyzed cellular transanimations.
%\newblock In A.F. Round, editor, \emph{Advances in Enzymology}, volume~2, pages
%  125--247. Academic Press, New York, 3rd edition, 1954.

%\bibitem[Keohane(1958)]{Keo:58}
%\newblock \emph{Power and Interdependence: World Politics in Transitions}.
%\newblock Little, Brown \& Co., Boston, 1958.

%\bibitem[Powers(1985)]{Pow:85}
%\newblock Is there a way out?
%\newblock \emph{Harpers}, pages 35--47, June 1985.

%\bibitem[Soukhanov(1992)]{Heritage:92}
%\newblock \emph{{The American Heritage. Dictionary of the American Language}}.
%\newblock Houghton Mifflin Company, 1992.

%\end{thebibliography}

\end{document}